# A Bayesian Approach for In-Situ Stress Prediction and Uncertainty Quantification for Subsurface Engineering

Ting Bao and Jeff Burghardt

Pacific Northwest National Laboratory, 902 Battelle Blvd, Richland, WA 99354.

**Abstract:** Many subsurface engineering applications require accurate knowledge of the in-situ state of stress for their safe design and operation. Existing methods to meet this need primarily include field measurements for estimating one or more of the principal stresses from a borehole, or optimization methods for constructing a 3D geomechanical model in terms of geophysical measurements. These methods, however, often contain considerable uncertainty in estimating the state of stress. In this paper, we build on a Bayesian approach to quantify uncertainty in stress estimations for subsurface engineering applications. This approach can provide an estimate of the 3D distribution of stress throughout the volume of interest and provide an estimate of the uncertainty arising from the stress measurement, the rheology parameters, and a paucity of measurements. The value of this approach is demonstrated using stress measurements from the In Salah carbon storage site, which was one of the first world's industrial carbon capture and storage projects. This demonstration shows the application of this Bayesian approach for estimating the initial state of stress for In Salah and quantifying the uncertainty in the estimated stress. Also, an assessment of a maximum injection pressure to prevent geomechanical risks from $CO_2$ injection pressures is provided in terms of the probability distribution of the minimum principal stress quantified by the approach. With the In Salah case study, this paper demonstrates that using the Bayesian approach can provide additional insights for site explorations and/or project operations to make informed-site decisions for subsurface engineering applications.

| Symbol list | | | |
|---|---|---|---|
| $x$ | List of uncertain parameters | $P_p$ | Effective pore pressure |
| $\sigma_h$ | Total minimum principal stress | $\rho_r$ | Rock density |
| $\sigma_H$ | Total maximum principal stress | $\delta$ | Kronecker delta |
| $\sigma_v$ | Total vertical stress | $\alpha$ | Biot's coefficient |
| $\varepsilon_h$ | Minimum horizontal strain | $\mathbf{C}$ | Stiffness tensor |
| $\varepsilon_H$ | Maximum horizontal strain | $E$ | Young's modulus |
| $\boldsymbol{\sigma}$ | Total stress tensor | $\nu$ | Poisson's ratio |
| $\boldsymbol{\varepsilon}$ | Strain tensor | $D$ | Given information for $x$ |
| $L_l$ | Lower bound | $\theta$ | Mean |
| $L_u$ | Upper bound | $\xi$ | Standard deviation |
| $\mu$ | Friction coefficient | | |

## 1 Introduction

The in-situ state of stress in the subsurface is important for the design and safe operation of many subsurface engineering projects. Examples of such applications are the design of hydraulic fracturing and borehole stability analysis in the oil and gas industry (Djurhuus and Aadnøy 2003; Lee and Ong 2018; Peška and Zoback 1995), and the safety and reliability design of rock engineering in the civil and mining industry (Martin et al. 2003; Stephansson and Zang 2012). Another example is geologic carbon storage (GCS), which requires knowledge of the in-situ state of stress for safe operation over decades. Injecting $CO_2$ in the subsurface can generate significant changes in stress, strain, and pore pressure in both the $CO_2$ storage reservoir and surrounding formations (Bao et al. 2021b; Burghardt 2017; White and Foxall 2016). The associated increase in pore pressure alters the initial state of stress over the course of a GCS project. This potentially could reactivate faults or create unintended fractures in the storage reservoir and sealing formations, resulting in geomechanical risks related to unacceptable levels of seismicity and $CO_2$ leakage (Bao et al. 2021a; Burghardt 2017; Zoback and Gorelick 2015).

Field measurements can help gain knowledge of the in-situ state of stress for subsurface engineering applications. Several methods currently exist for estimating one or more of the principal stresses from measurements in boreholes (Zoback 2010). These methods, in most cases, are restricted to providing information about the stress in the immediate vicinity of the borehole, and can take a considerable amount of time and equipment to execute. Therefore, the number of such tests in any given project is often limited. This leaves most of the rock volume at most sites uncharacterized in terms of direct stress measurements. Given these challenges, optimization methods that integrate with either numerical or analytical models have been developed to interpolate and extrapolate stress measurements into a 3D volume of interest, e.g., Stavropoulou et al. (2007) and Zhao et al. (2012). Generally, these methods use geologic models and geophysical measurements to construct a 3D distribution of rheology model parameters (elastic, viscoelastic, plastic, etc.). Then the 3D boundary value problem is solved with the lateral boundary conditions being fitting parameters that are chosen, such that the distribution of stress provides the best fit with measured values. It is well recognized that direct measurement methods are very helpful but yield a significant degree of uncertainty, and in many cases can only provide relatively wide limits on the possible value of the principal stresses and their directions. Also, most of the optimization methods to fit geophysical measurements are deterministic, that is, only provide the best-fit model to match the most likely stress state from measured data and do not quantify any uncertainty in geophysical measurements and modeling parameters.

A method for both in-situ stress estimation and uncertainty quantification is therefore highly desirable for a wide range of subsurface engineering applications. In particular, applications needing to make critical decisions based on the in-situ stress state would benefit greatly from such a method. GCS and enhanced geothermal systems (EGS) are two examples of this application type. For both of these applications, the geomechanical risk of induced seismicity is one of the primary concerns of safe GCS operations. This risk is determined largely by the initial state of rock stress and how this stress state evolves with time during and after injection. To help meet this need of accounting for uncertainty in stress estimation for operational decisions, the State-of-Stress Analysis Tool (SOSAT) has been developed (NRAP 2021). This tool uses a Bayesian approach to quantify the uncertainty in the in-situ state of stress and in how the state of stress will evolve as a result of $CO_2$ injection (Burghardt 2018). Using a known range of pore pressures, SOSAT could estimate and provide the likelihood of fault reactivation and fracture generation in terms of the in-situ state of stress.

The merit of SOSAT thus makes it useful for geomechanical analyses to make informed decisions for currently active GCS projects, and for, in hindsight, evaluating geomechanical risks of operational decisions for past GCS projects. In this study, we use stress measurement data from the In Salah GCS project, which was one of the pioneering industrial-scale carbon capture and storage projects in the world. Measured stress data available from this project can be used for demonstrating the value of the Bayesian approach used in SOSAT. Briefly, the In Salah GCS project injected over 3.8 million tons of $CO_2$ between 2004 and 2011 into a ~20 m-thick sandstone reservoir. The reservoir is overlaid by ~950 m-thick caprock formations with varying levels of permeability and porosity. During the GCS operation, monitoring data in June 2011 indicated that $CO_2$ probably migrated up into the lower portion of the caprock (White et al. 2014). The InSAR (Interferometric Synthetic Aperture Radar) observations also showed significant uplift of the ground deformation above all three $CO_2$ injection wells, with the deformation around well KB-502 exhibiting an unusual double-lobe pattern (see Fig. 1c). Numerous studies have tried to determine the cause of these observations (Bjørnarå et al. 2018; Morris et al. 2011; Rinaldi and Rutqvist 2013; Rutqvist et al. 2010; Shi et al. 2012; Vasco et al. 2010). For example, Shi et al. (2012) conducted the pressure history matching analysis and concluded that the double-lobe pattern around KB-502 is caused by tensile opening of a non-sealing fault in the caprock. White et al. (2014) comprehensively reviewed all hypotheses proposed in the literature and argued that hydraulic fracturing caused by high injection pressures is responsible for the double-lobe pattern. Though these studies have provided valuable insights, the interpretations are based on observations (e.g., observed InSAR deformations, seismic velocity anomalies, and injection behavior) that are associated with considerable uncertainty; and the results based on deterministic methods (e.g., best-fit models) are just one realization, which fails to quantify any uncertainty in either the field measurements or the modeling parameters.

The SOSAT approach, as emphasized above, could address these limitations to offer an improved geomechanical analysis. In this paper, the objective is to build on a Bayesian approach to construct the in-situ state of stress for the In Salah GCS site and quantify the uncertainty arising from the stress measurement, the rheology parameters, and a paucity of measurements. Lecampion and Lei (2010) also used a Bayesian method to calibrate a 3D geomechanical model for the In Salah site. The current paper will make contributions that advance the approach and analysis of Lecampion and Lei (2010) in the following three ways. First, this paper will include uncertainty in rock rheology in the analysis, whereas Lecampion and Lei (2010) took rock rheology as deterministically known. Second, this paper will show and analyze two horizontal principal stresses and uncertainty at locations having no stress measurements. Third, this paper will demonstrate the results in terms of the probability that the excessive injection pressure caused hydraulic fracturing of the lower caprock In Salah GCS.

This paper is organized as follows. In Section 2, a Bayesian approach for both 1D elastic-tectonic and 3D poroelastic models is presented. Section 3 presents the results of the application of these two models to the In Salah site including a convergence analysis. A discussion regarding insights and recommendations provided in the Bayesian framework for In Salah GCS is presented in Section 4. Finally, conclusions and recommendations for future efforts are summarized.

## 2 Theory and Method
### 2.1. Bayesian method and Uncertainty

Uncertainty in subsurface engineering arises from many sources, such as spatial and temporal variability of the material properties in a natural 3D volume, measurement errors of the properties/variables, selection of modeling approaches to reflect true physics, and approximation of physical modeling parameters to estimate variables of interest (Miranda et al. 2009). For GCS, stress measurements are often made at sites early in the design stage to characterize the initial stress field. With a set of discrete stress measurements and a geomechanical model for the behavior of the system, the uncertainty to be quantified primarily arises from stress measurement errors, the choice of modeling parameters, and everywhere that no stress measurements are collected. Among the modeling parameters that must be chosen, the most critical includes the form of the constitutive model and the associated parameters for each spatial position in the system being modeled, as well as the boundary conditions that approximate prevailing tectonic processes. The outputs of such a model can then include the initial state of stress (often called a static model) and the change in stress induced by project operations (often called a dynamic model). For the purposes of the present study, we will assume an *a priori* elastic response and spatially uniform properties within each formation. Implications for relaxing these two assumptions to better fit realistic heterogeneity and complex rock rheology will certainly deserve an evaluation, this however is beyond the scope of this paper.

A Bayesian inversion approach is used for estimations of in-situ stress with uncertainty quantification using Bayes Law (Bayes 1763)

$$p(\boldsymbol{x}|\boldsymbol{D}) = \frac{p(\boldsymbol{D}|\boldsymbol{x})p(\boldsymbol{x})}{p(\boldsymbol{D})} \tag{1}$$

where $x$ is a list of uncertain parameters whose probability distribution will be sought (e.g., two horizontal (principal) in-situ stresses $\sigma_h$ and $\sigma_H$, elastic parameters $E$ and $v$, and horizontal tectonic strains $\varepsilon_H$ and $\varepsilon_h$, etc.); $D$ refers to the available observations being used to update knowledge of the uncertain parameters; $p(D|x)$ is the likelihood function; $p(x)$ is the prior distribution; and $p(x|D)$ is the posterior distribution. This Bayesian inversion analysis is unlike numerical optimization approaches commonly used to solve inversion problems only finding best-fit values for modeling parameters, the Bayesian approach seeks to compute the posterior probability distribution for the uncertain parameters in light of the measured data. Measured data are treated as observed stochastic variables in this Bayesian analysis.

In GCS, observed stochastic variables are usually $\sigma_h$ and $\sigma_H$ that will be used for evaluation of fault failure and fracture re-activation/generation. Techniques such as hydraulic fracture-based stress tests (mini-frac, extended leak-off tests (XLOT), diagnostic fracture injection tests (DFIT), etc.) are used to provide the magnitude of $\sigma_h$. Observations from borehole image logs can be used to determine the stress directions by observing the orientations of induced fractures and borehole breakouts. Breakouts and drilling-induced tensile fractures (DIFT) can also be used (somewhat controversially) to estimate the magnitude of $\sigma_H$. The Bayesian inversion modeling thus can be unitized to compute the posterior probability of $\sigma_h$ and $\sigma_H$ via either the deterministic 1D tectonic-elastic model (Thiercelin and Plumb 1994) along a 1D depth profile, given by

$$\sigma_h = \frac{v}{1-v}(\sigma_v - \alpha P_p) + \frac{E}{1-v^2}\varepsilon_h + \frac{Ev}{1-v^2}\varepsilon_H + \alpha P_p \quad (2)$$

$$\sigma_H = \frac{v}{1-v}(\sigma_v - \alpha P_p) + \frac{E}{1-v^2}\varepsilon_H + \frac{Ev}{1-v^2}\varepsilon_h + \alpha P_p \quad (3)$$

or the deterministic 3D model of linear poroelasticity in a 3D volume, expressed by

$$\nabla \cdot \boldsymbol{\sigma} + \rho_r \mathbf{g} = 0 \quad (4)$$

$$\boldsymbol{\sigma} = \mathbf{C} : \boldsymbol{\varepsilon} + \alpha P_p \boldsymbol{\delta} \quad (5)$$

where $\boldsymbol{\sigma}$ is the total stress tensor, $P_p$ is the effective pore pressure, $\rho_r$ is the rock density, $\boldsymbol{\delta}$ is the Kronecker delta, $\boldsymbol{\varepsilon}$ is the infinitesimal strain tensor, $\alpha$ is the Biot's coefficient [=0.6 in this study (Burghardt 2018)], and **C** is the fourth-order stiffness tensor involving with $E$ and $v$. Here and throughout strain and stress are both taken to be positive in compression.

The 1D model given by Eqs. (2-3) assumes that $\varepsilon_H$ and $\varepsilon_h$ are uniform along depth and elastic parameters are uniform for each formation. The third principal stress $\sigma_v$ is oriented vertically and computed in terms of the weight of overlying rocks at any depth. Compared to this 1D model, the 3D model is more computationally costly, but can account for geological features such as faults, heterogeneity, and 3D structural features such as anticlines, synclines, etc. As an initial evaluation, this paper does not consider such more generally 3D heterogeneity but will consider isotropic linear elastic materials and uniform rock properties within each formation.

Bayesian inversion modeling requires first to specify prior distributions for modeling parameters as their historical or expert information before any new measured data are collected. This is a quantitative way to embedding *a priori* assumptions into the analysis. In this study, we assume a uniform probability value for all compressive stresses according to Burghardt (2018) and assign uniform distributions for modeling parameters in $x$ via

$$E, v, \varepsilon_H, \varepsilon_h \sim \text{Uniform}\{L_l, L_u\} \tag{6}$$

where $L_l$ and $L_u$ are the lower and upper bounds. In general, $v$ for rocks is between 0 and 0.5 and $E$ is in a range of 1 and 100 GPa, so these ranges are used to define the prior models for these parameters. To compute $\sigma_h$ and $\sigma_H$ during the joint Bayesian analysis, $\varepsilon_H$ and $\varepsilon_h$ are specified conservatively within -10~10 (note that in the 3D case, horizontal displacements are specified).

The analysis can be significantly improved if more informative prior models can be used. For example, if one of the formations of interest was known to be an igneous basement formation, then the lower bound for the prior model of $E$ could be significantly increased, which would accelerate model convergence and accuracy. In contrast, overly prescriptive prior models can cause slow convergence or, even worse, convergence to incorrect results. Therefore, it is generally best to err on the side of less informative prior distributions and let the model learn from real observations.

### 2.2. Likelihood function for field observations and stress constraints

The likelihood function for field stress measurements is constructed using a normal distribution

$$\sigma_H, \sigma_h \sim \text{Normal}(\theta, \xi^2) \qquad (7)$$

where $\theta$ and $\xi$ are the mean and standard deviation for stress. This likelihood function represents the probability of observing field stresses that given modeling parameters are trying to reproduce.

In addition to observations from well tests, it is also helpful to add constraints on the state of stress based on frictional faulting arguments and regional observations. The first constraint for frictional faulting arguments from a physical perspective is to constrain $\sigma_h$ and $\sigma_H$ within their upper and lower bounds. For example, $\sigma_h \leq \sigma_H$ may be true for any case, and the upper bound of $\sigma_H$ at any given depth and pore pressure should be constrained via the friction coefficient $\mu$ according to the frictional strength of faults (Jaeger et al. 2009; Zoback et al. 2003) via

$$\begin{cases} \text{SS: } \sigma_H \leq (\sigma_h - P_p)\left(\sqrt{\mu^2+1}+\mu\right)^2 + P_p \\ \text{TF: } \sigma_H \leq (\sigma_v - P_p)\left(\sqrt{\mu^2+1}+\mu\right)^2 + P_p \\ \text{NF: } \sigma_v \leq (\sigma_h - P_p)\left(\sqrt{\mu^2+1}+\mu\right)^2 + P_p \end{cases} \qquad (8)$$

where NF=normal-faulting, SS=strike-slip faulting, and TF=thrust-faulting. Eq. (8) uses the frictional strength of faults and fractures present in a rock mass at any size to limit the possible range of $\sigma_H$ or $\sigma_v$ magnitudes in the subsurface stress state generally. This constraint in general applies at any site without any observations. To be conservative, $\mu = 1$ is adopted here to limit the maximum magnitude of $\sigma_H$ or $\sigma_v$.

Another constraint on stress in terms of regional observations is whether the initial stress regime at a site is the NF, SS, or TF condition. This determines which stress is the minimum or maximum among three principal stresses. There are many methods available in the literature for the setup of stress constraints and a simple way is to utilize the stress regime factor $Q$ (Pistre et al. 2009) to constrain $\sigma_H$ and $\sigma_h$, given by

$$Q = \begin{cases} \text{NF}: \dfrac{\sigma_H - \sigma_h}{\sigma_v - \sigma_h} \\ \text{SS}: 2 - \dfrac{\sigma_v - \sigma_h}{\sigma_H - \sigma_h} \\ \text{TF}: 2 + \dfrac{\sigma_h - \sigma_v}{\sigma_H - \sigma_v} \end{cases} \qquad (9)$$

The SS stress regime, for example, needs to have $1 < Q \leq 2$. In later sections of this paper, we will show how this stress regime constraint impacts the stress estimation and uncertainty associated with in the analysis.

## 3  Application of Bayesian Approach

This section demonstrates the application of Bayesian inversion modeling for a real GCS case to estimations of the in-situ state of stress with uncertainty quantification. The CGS case focused on here is the In Salah CGS site earlier mentioned in the Introduction. $CO_2$ was injected into an anticlinal structure of the sandstone reservoir through injection wells KB-501, KB-502, and KB-503 (Fig. 1a). In Table 1, leak-off test (LOT) and formation integrity test (FIT) data for this CGS site is available at different depths measured along three $CO_2$ injection wells and also several gas production wells at the site (Lecampion and Lei 2010). This type of data estimates $\sigma_h$ at different spatial locations, with an approximate standard deviation of 3 MPa to represent the uncertainty inherent in the interpretation of test results.

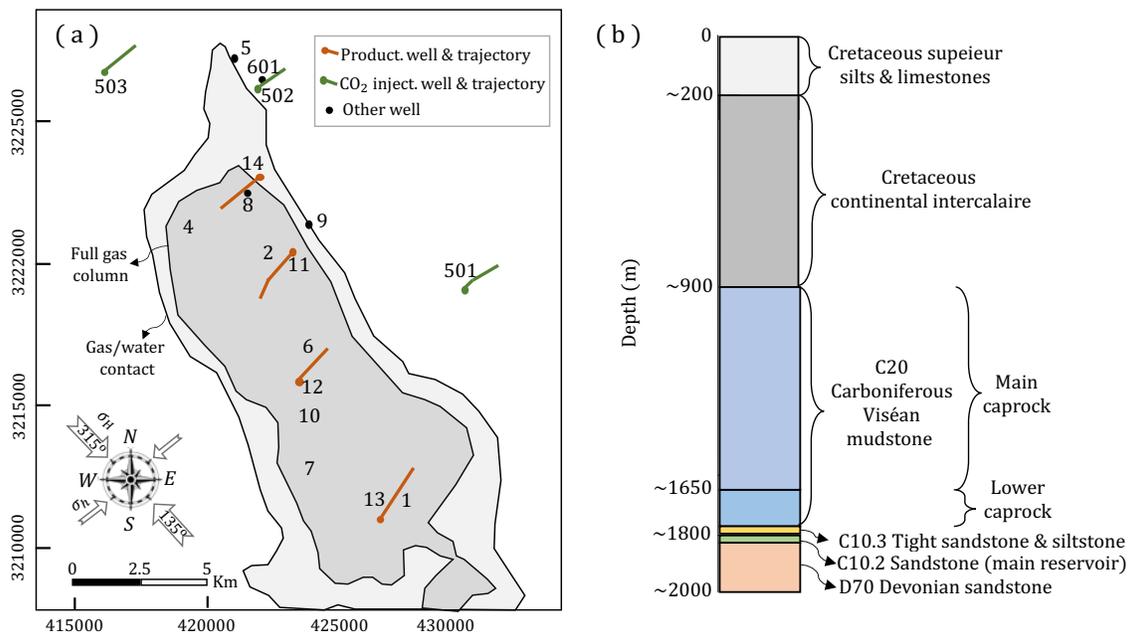

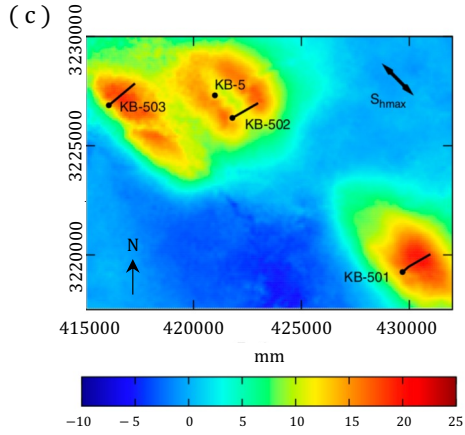

**Fig. 1.** (a) In Salah GCS site map to show the $CO_2$ injection and gas production wells and well locations [after Shi et al. (2012) ], (b) stratigraphic column along vertical depth for $CO_2$ injection well KB-502 [approximated depths according to Bjørnarå et al. (2018)], and (c) Double-lode pattern of ground surface deformations measured by InSAR [adapted from White et al. (2014)].

The orientations of $\sigma_h$ and $\sigma_H$ were characterized to be ~N45E and ~N135S, respectively (Fig. 1a). For KB-502, the double-lobe pattern observed in the ground surface deformations can be seen in Fig. 1c. The stratigraphic column is shown in Fig. 1b, consisting of a ~20 m-thick C10.2 main sandstone reservoir for $CO_2$ storage. A ~20m-thick C10.3 tight sandstone/siltstone formation sits above C10.2. The C20 caprock above C10.3 consists of low-permeability rocks and is grouped into the lower caprock and main caprock to seal the vertical migration of $CO_2$. Table 1 also shows LOT and FIT data for KB-502 only. This data is obtained from Shi et al. (2012) and provides a high resolution of measured $\sigma_h$ that we will use later in the 1D model for a geomechanical analysis.

**Table 1.** LOT and FIT data for $\sigma_h$ [MC=main caprock, LC=lower capcok]

| Measured data used for 3D | | | | Measured data used for 1D | | | |
|---|---|---|---|---|---|---|---|
| Well ID | Depth (m) | $\sigma_h$ (MPa) std. dev.=3 | Formation | Well ID | Depth (m) | $\sigma_h$ (MPa) std. dev.=3 | Formation |
| KB-502 | 1298 | 17.57 | MC | KB-502 | 1298 | 17.57 | MC |
| KB-12 | 1376 | 17.60 | | | 1436 | 19.13 | |
| KB-14 | 1423 | 19.00 | | | 1506 | 20.32 | |
| KB-502 | 1498 | 20.10 | | | 1616 | 21.40 | |
| KB-503 | 1542 | 20.30 | | | 1657 | 26.20 | LC D1 |
| KB-12 | 1580 | 24.60 | | | 1671 | 24.64 | LC D2 |
| KB-11 | 1648 | 26.10 | | | 1706 | 27.10 | LC D3 |
| KB-14 | 1666 | 24.50 | LC D1 | | 1782 | 29.26 | Tight C10.3 |
| KB-502 | 1734 | 27.00 | LC D2 | | 1851 | 22.84 | Sandstone D70 |
| KB-501 | 1766 | 29.30 | LC D3 | | | | |
| KB-502 | 1782 | 29.26 | Tight C10.3 | | | - | |
| KB-502 | 1851 | 22.84 | Sandstone D70 | | | | |

Shown in Table 2 is the workflow of Bayesian inversion modeling proposed in this paper. Depending on the need, a deterministic model, either the 1D model via Eqs. (2-3) or 3D model via Eqs. (4-5), can be employed to compute $\sigma_h$ during the Bayesian inversion analysis. In Step 2, the needed variables are, for example, the depth of stress measurements, $P_p$, and the likelihood function for measured $\sigma_h$. The distribution of $P_p$ is assumed to be linear here as did in the existing studies (Lecampion and Lei 2010; Rutqvist et al. 2010). This distribution assumption, however, may not reflect the real pore pressure distribution because of the gas-induced overpressure in the reservoir and caprock. Thus, the analysis here is treated as a first-order estimation.

**Table 2.** Workflow of Bayesian inversion modeling

| | |
|---|---|
| Step 1 | Construct priors for modeling parameters via Eqs. (6-7) |
| Step 2 | Construct needed variables for $\sigma_h$ calculation and compute $\sigma_h$ via deterministic 1D or 3D models (if necessary, activate stress constraints) |
| Step 3 | Propose probability distributions for modeling parameters and start sampling via MCMC |
| Step 4 | Output and store MCMC trace for all parameters |

Computing posterior distributions here is achieved using the Markov chain Monte Carlo (MCMC) simulation. The MCMC simulation draws a sequence of sample values with the distribution of the sampled draws, depending only on the last value. The MCMC Metropolis-Hastings algorithm (Hastings 1970) is employed to draw a sequence of random samples from the posterior distribution. This is performed in Python using PyMC (Patil et al. 2010), a Python module for Bayesian stochastic modeling. The detailed theory of the MCMC simulation and its algorithm for posterior sampling are not repeated here as they are well and comprehensively described in the literature, e.g., Brooks et al. (2011).

In this paper, we perform the 1D and 3D Bayesian inversion analyses separately for the In Salah GCS site. The 3D analysis here can make full use of the $\sigma_h$ measurements from different wells listed in Table 1, and account for the geologic model information (i.e., structural topography). Since the In Salah GCS filed has an anticlinal structure (Mathieson et al. 2009), using its real geologic model for 3D geomechanics analysis could be the best option to match the stress measurements. To the authors' knowledge, this geological model is not publicly available and our requests for the structural topography from the project's governing body were not granted. We thus approximate it by synthetically generating a roughly symmetrical anticline structure (with a maximum angle of about 10 degrees between the limb and horizontal plane on the *x-z* plane, see Fig. A1b in Appendix). The 1D analysis, however, can take advantage of computational efficiency and high resolution of depth along a well to quickly obtain simulation results. When the rock formation is assumed to be homogenous and has a relatively flat structure, the 1D analysis could provide reasonable estimations for the 3D analysis. Here, the 1D analysis for KB-502 is demonstrated using the measured data in Table 1 for this well. The results will be presented in the following.

## 4 Results

### 4.1 MCMC Convergence Analysis

The minimum number of samples in the MCMC analysis is very often difficult to know to guarantee a satisfactory approximation to the target posterior density. As a result, the convergence check is an essential step in the analysis to gain confidence. Several different approaches in the literature can be utilized for convergence check. Here, we first perform an informal evaluation of trace plots of all modeling parameters. As an example, Fig. 2 plots the traces of three modeling parameters and $\sigma_h$ for the LC D3 formation (one unit in the lower caprock listed in Table 1), where two Markov chains were performed in the MCMC analysis. Each of the chains ran with one million MCMC iterations and started from a randomly chosen, different starting point. As can be seen in Fig. 2, the traces from the chains are overlapped with each other. Fig. 2 also shows the tending of each chain toward the almost same posterior distributions. The traces for $E$ are nearly uniformly distributed with 1-100 GPa, indicating the uniform $E$ posterior distributions similar to its prior distributions for this example.

After the informal evaluation, the formal method, called the Gelman-Rubin statistic (Gelman and Rubin 1992), is employed to further diagnose MCMC convergence. Briefly, this method calculates both the between-chain variance ("B" variable) and within-chain variance ("W" variable) to examine whether the chains are very different to suggest a lack of convergence. A simple parameter used to reveal this is the $\hat{R}$ ratio. It is generally believed that the chains are converged when $\hat{R}$ approaches 1.0. The $\hat{R}$ values for all modeling parameters, $\sigma_h$, and $\sigma_H$ are found within 1.0-1.05 in the analysis, and Fig. 3 shows some examples of modeling parameters in different rock formation classes and $\sigma_h$. This further verifies reasonable convergence obtained in our Bayesian inversion analysis.

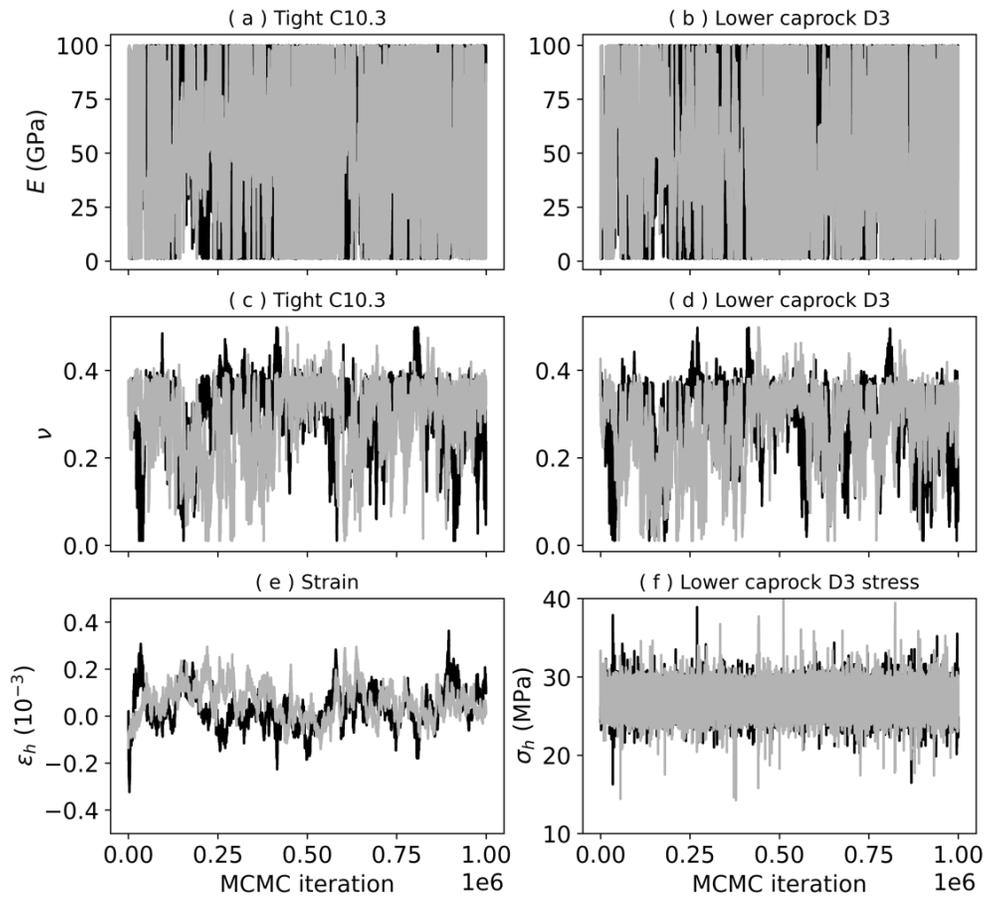

**Fig. 2.** Examples of MCMC trace plots of $E$, $\nu$, $\varepsilon_h$, and $\sigma_h$ using a million iterations ($\varepsilon_H$ not shown here but also a stochastic parameter). $E$ and $\nu$ here stand for the rock formation of Tight C10.3 and LC D3. $\sigma_h$ values predicted here are at 1706 m in LC D3 in Table 1.

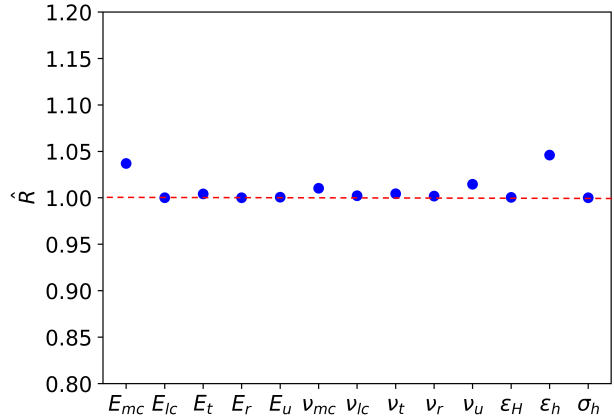

**Fig. 3.** Gelman-Rubin $\hat{R}$ values for some examples of modeling parameters and $\sigma_h$.

### 4.2 Use of 1D Model for KB-502 In-Situ Stress Prediction

In the 1D analysis, given that $\sigma_h$ measurements in Table 1 are available starting from the main caprock down to the underlying sandstone D70, we consider eight rock formation classes, consisting of the main caprock, four lower caprock units, tight C10.3, reservoir C10.2, and underlying D70. The lower caprock is refined into four units that are comprised of LC D1-3 in Table 1 with measured $\sigma_h$ data available and another one laying between LC D3 and tight C10.3 (termed LC bottom and marked by "N" in Fig. 5) without measured $\sigma_h$. This will allow us to analyze $\sigma_h$ uncertainty at locations lacking of stress measurements and also carefully look into the possibility of hydraulic fracturing at the bottom of the lower caprock under injection pressures.

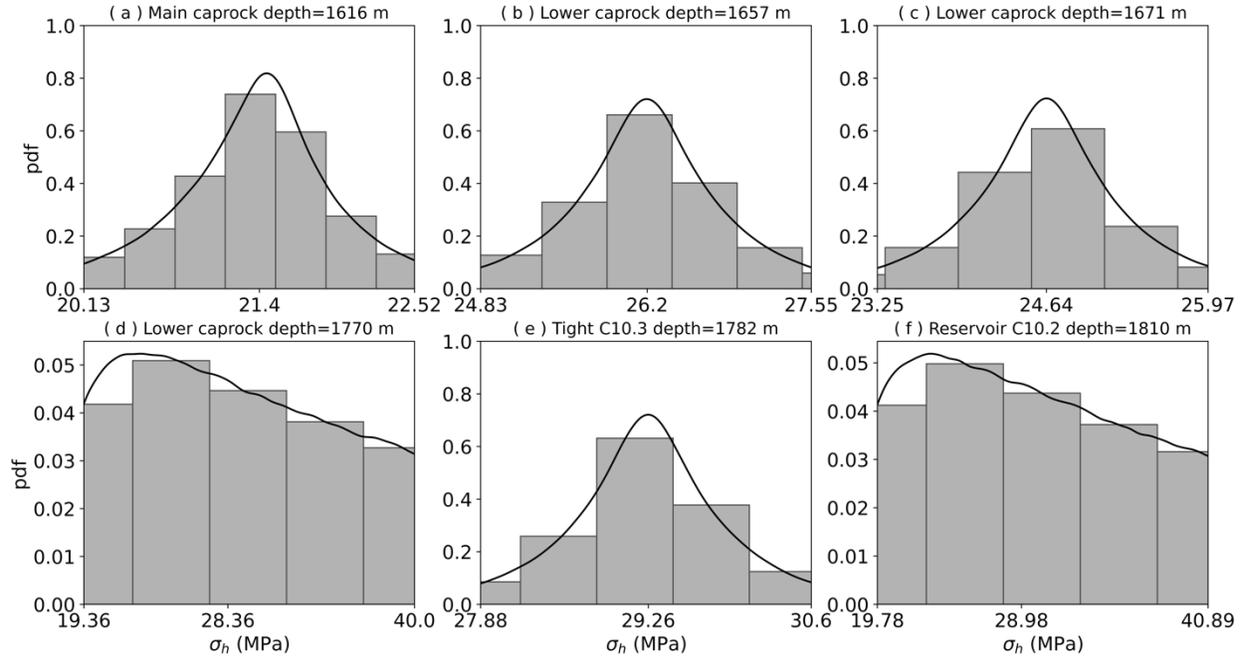

**Fig. 4.** Histograms and distributions of MCMC draws for $\sigma_h$ in the (a) main caprock, (b) LD D1, (c) LD D2, (d) LC bottom, (e) tight C10.3, and (e) reservoir C10.2. Probability density function=pdf.

Given that in the lower caprock, we have some depths with $\sigma_h$ measurements available but without in the LC bottom and reservoir C10.2, Fig. 4 plots both histograms and distributions of MCMC draws for $\sigma_h$ at six depths to cover the above two cases. The *x* tick marks denote the median and the lower and upper of 95% confidence interval (CI) for each predicted $\sigma_h$. At depths where we have $\sigma_h$ measurements, in Figs. 4a-c and 4e, we see small uncertainty in $\sigma_h$ and whose tolerance (both lower and upper) is less than 1.5 MPa. Lacking of $\sigma_h$ measurements results in large uncertainty in $\sigma_h$ with a tolerance of 9~12 MPa in the LC bottom (Fig. 4d) and reservoir C10.2 (Fig. 4f). This indicates the importance of adding informative priors, for example, for the elastic properties with a sonic log or core samples to reduce $\sigma_h$ uncertainty. This will be discussed further in Section 5.2.

Due to the stress constraints presented in Section 2.2 and large uncertainty in priors of the elastic parameters, $\sigma_h$ in Fig. 4d and Fig. 4f exhibits a skewed distribution-like shape rather than a normal distribution-like shape. We thus here and throughout adopt the median to approximate the most possible value of $\sigma_h$ (note that the median and mean for a normal distribution is identical). The $\sigma_h$ median in the LC bottom in Fig. 4e is higher than those in the lower caprock sitting above the LC bottom, but the reservoir C10.2 in Fig. 4f has a $\sigma_h$ median slightly smaller than the tight C10.3. This means that the reservoir C10.2 has a higher likelihood of hydraulic fracturing than the tight C10.3.

**Table 3.** Posterior modeling elastic parameters and $\sigma_h$ for KB-502. Lower and Upper refer to the lower and upper of 95% confidence interval (CI).

| Formation | $\sigma_h$ (MPa) std. dev.=3 | MCMC 95% CI |||||||||
|---|---|---|---|---|---|---|---|---|---|---|
| | | *E* (GPa) ||| *v* ||| $\sigma_h$ (MPa) |||
| | | Lower | Median | Upper | Lower | Median | Upper | Lower | Median | Upper |
| MC | 17.57 | 5.86 | 49.32 | 94.53 | 0.11 | 0.26 | 0.30 | 16.51 | 17.63 | 18.87 |
| | 19.13 | | | | | | | 18.15 | 19.27 | 20.38 |
| | 20.32 | | | | | | | 18.95 | 20.10 | 21.20 |
| | 21.40 | | | | | | | 20.13 | 21.40 | 22.52 |
| LC D1 | 26.20 | 5.15 | 41.32 | 91.66 | 0.09 | 0.29 | 0.36 | 24.83 | 26.20 | 27.55 |
| LC D2 | 24.64 | 3.24 | 24.72 | 84.88 | 0.08 | 0.27 | 0.34 | 23.25 | 24.64 | 25.97 |
| LC D3 | 27.10 | 5.32 | 47.33 | 94.11 | 0.10 | 0.29 | 0.36 | 25.71 | 27.09 | 28.44 |
| C10.3 | 29.26 | 6.19 | 50.48 | 94.69 | 0.12 | 0.30 | 0.37 | 27.88 | 29.26 | 30.60 |
| D70 | 22.84 | 6.06 | 52.94 | 94.52 | 0.06 | 0.21 | 0.27 | 21.48 | 22.84 | 24.20 |

Table 3 shows the posterior $\sigma_h$ at all measurement depths and the corresponding posterior elastic parameters *E* and *v*, as well as $\sigma_h$ measurements for comparison. It is seen that the $\sigma_h$ medians well match the measured $\sigma_h$ data. Deviated from the median, the lower and upper are quantified to be a maximum value of about 1.39 and 1.36 MPa, respectively. So, the uncertainty in MCMC $\sigma_h$ estimations is found to be approximately 1.1~1.4 MPa (maximum lower and upper) deviated from the medians.

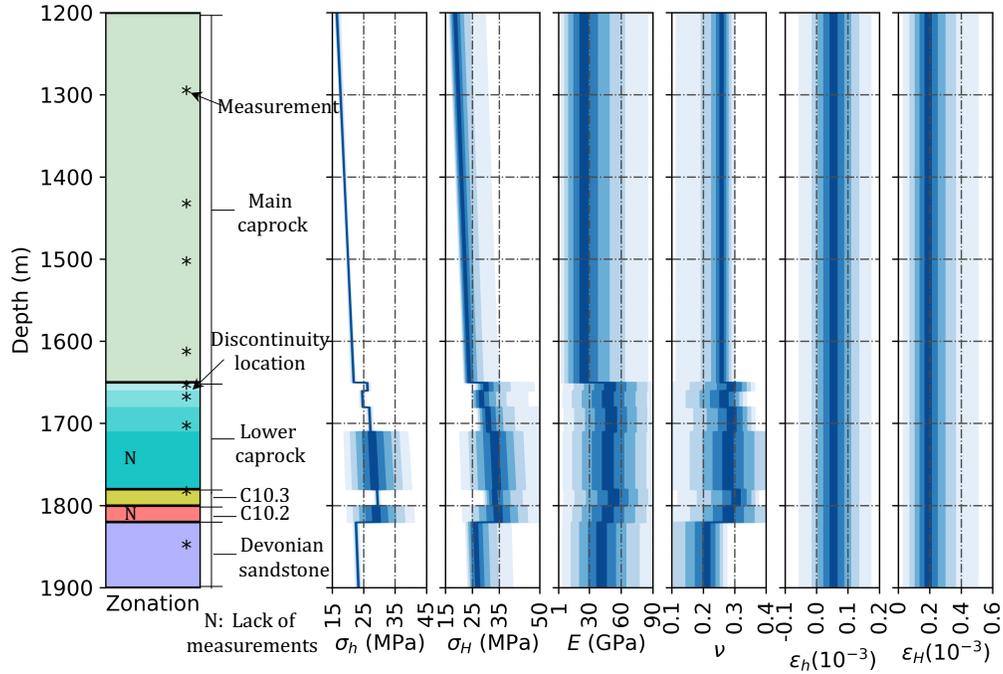

**Fig. 5.** Posterior $\sigma_h$, $\sigma_H$, $E$, $v$, $\varepsilon_H$ and $\varepsilon_h$ for KB-502 along 1D depth from 1200m to 1900m.

One of the key barriers to ensuring safe operations of, and reducing the geomechanical risks associated with GCS is the high degree of uncertainty in the down-hole stress measurements, particularly from locations that lack direct stress measurements. However, we have inferred from Table 3 that the posterior $\sigma_h$ has small uncertainty at all measurement depths (1.1~1.4 MPa). Next, we can look into other locations where there are no direct stress measurements available. Using the available trace data for $E$, $v$, $\varepsilon_H$ and $\varepsilon_h$ from the MCMC draws, we are able to calculate $\sigma_h$ and $\sigma_H$ at any location via Eqs. (2-3). Fig. 5 shows the posterior $\sigma_h$, $\sigma_H$, $E$, $v$, $\varepsilon_H$ and $\varepsilon_h$ for KB-502 along 1D depth of 1200-1900 m that covers rock formation classes having measured $\sigma_h$ and two classes without stress measurements. These two rock classes in Fig. 5 are the LC bottom and reservoir C10.2 (marked by "N"). The stress measurements are marked by the star symbol and their values are detailed in Table 3. As shown in Fig. 5, the lack of stress measurements in the LC bottom and reservoir C10.2 leads to significant and larger uncertainty in $\sigma_h$ than other locations. For example, 95% CI for $\sigma_h$ is about 20~40 MPa in both the LC bottom and reservoir C10.2.

For $\sigma_H$ in Fig. 5, its uncertainty is found to be significant along 1200-1900 m, even at locations where we have $\sigma_h$ measurements, because of no measurements to constrain $\sigma_H$. At any location, the median of $\sigma_H$ is not less than that of $\sigma_h$, and given that $\sigma_v$ has the highest value among three in this case, so the posterior stress result here yields the initial stress regime in the NF condition. This could be constrained to other initial stress regime conditions and will be further discussed later in Section 5.1.

## 4.3 Use of 3D Model for In-Situ Stress Prediction for In Salah

The 3D Bayesian modeling was performed using finite element methods for a model whose dimension is $20\text{km} \times 10\text{km} \times 620\text{m}$ ($x \times y \times z$). 14,336 hexahedral elements are utilized (see Fig. A1a in Appendix) with an element number of 32 in the *x* and *y* directions and 14 in the *z* direction, respectively. Uniform lateral displacements are assigned on four lateral boundaries. A linear pore pressure distribution is employed here and the pore pressure is assumed to be 17.9 MPa at 1800 m according to Rutqvist et al. (2010). In this 3D analysis, seven rock formation classes in total are considered, i.e., the reservoir C10.2 plus the other six listed in Table 1. Since running 3D Bayesian modeling is very computationally expensive, two thousand MCMC iterations were run for this 3D analysis to save the computational cost. To get reasonable convergence, the *MAP* function in PyMC was performed first to obtain an estimate of the most probable *x* values (e.g., tectonic displacements and elastic parameters) while maximizing the posterior pdf for $\sigma_h$. This step is equivalent to a general inversion analysis by minimizing an objective function via optimization techniques. The estimated *x* values were then chosen as the starting points to run the MCMC simulation. This will allow the MCMC simulation to start from the most probable *x* values. Because a relatively small number of MCMC iterations and uniform lateral displacements are used here, this 3D analysis is regarded as a first order estimation.

The posterior $\sigma_h$ and measured data for multiple wells can be seen in Table 4. In general, the median of posterior $\sigma_h$, which is a primary interest because it is usually used in deterministic methods as the mean stress state, agrees well with the measurements from different wells. There is a slight difference between them and a relatively large mismatch appears at 1580 m for KB-12.

This is likely caused by the fact that the synthetic anticlinal structure of the geologic model used here is different from the realistic geologic model.

**Table 4.** Posterior $\sigma_h$ of the 3D modeling case vs measured $\sigma_h$.

| Well ID | Depth (m) | Measurement $\sigma_h$ (MPa) std. dev.=3 | MCMC 95% CI for $\sigma_h$ (MPa) | | |
|---|---|---|---|---|---|
| | | | Lower | Median | Upper |
| KB502 | 1298 | 17.57 | 16.08 | 16.92 | 17.65 |
| KB12  | 1376 | 17.60 | 16.69 | 17.57 | 18.32 |
| KB14  | 1423 | 19.00 | 18.00 | 18.98 | 19.75 |
| KB502 | 1498 | 20.10 | 18.11 | 19.03 | 19.81 |
| KB503 | 1542 | 20.30 | 18.76 | 19.75 | 20.55 |
| KB12  | 1580 | 24.60 | 19.38 | 20.36 | 21.18 |
| KB11  | 1648 | 26.10 | 24.17 | 26.50 | 28.61 |
| KB14  | 1666 | 24.50 | 22.19 | 24.11 | 25.21 |
| KB502 | 1734 | 27.00 | 27.04 | 29.01 | 30.70 |
| KB501 | 1766 | 29.30 | 24.83 | 27.29 | 29.01 |
| KB502 | 1782 | 29.26 | 24.82 | 27.28 | 29.00 |
| KB502 | 1851 | 22.84 | 19.76 | 21.37 | 22.77 |

To visualize the $\sigma_h$ profile, we plot the 2D slices of $\sigma_h$ in Fig. 6 for KB-502 as an example. These two slices are in the directions of $\sigma_h$ (y axis) and $\sigma_H$ (x axis). The entire $\sigma_h$ distribution in terms of its posterior median can be seen in Fig. A1c in Appendix. We see in Fig. 6 that $\sigma_h$ is not linearly increasing with depth, which is similar to the results observed in Section 4.2 for the 1D analysis (Fig. 5). For the region in the lower caprock, C10.3, and C10.2, the median $\sigma_h$ is approximately lower than 30 MPa.

Fig. 7 plots the vertical profile of $\sigma_h$ and $\sigma_H$ for KB-501. This is an example of showing two horizontal stresses where no or a few $\sigma_h$ measurements are available for a well at the site. The black line is the median of $\sigma_h$ from MCMC draws and the grey shading denotes 95% CI of MCMC $\sigma_h$ estimates. Because of relatively more measurements above 1650 m from all wells and fewer measurements below that depth, $\sigma_h$ in the region of the main caprock has smaller uncertainty than the region in the lower caprock, C10.3, and C10.2 (Fig. 7a). In those two discussed regions, the uncertainty in $\sigma_H$ is more significant than $\sigma_h$ (Fig. 7b). However, because no measured $\sigma_h$ is provided in C10.2 (depth=1800~1820 m for KB502), this leads to significant uncertainty in both $\sigma_h$ and $\sigma_H$ in C10.2 for KB-501. These are similar to the observations in the 1D analysis.

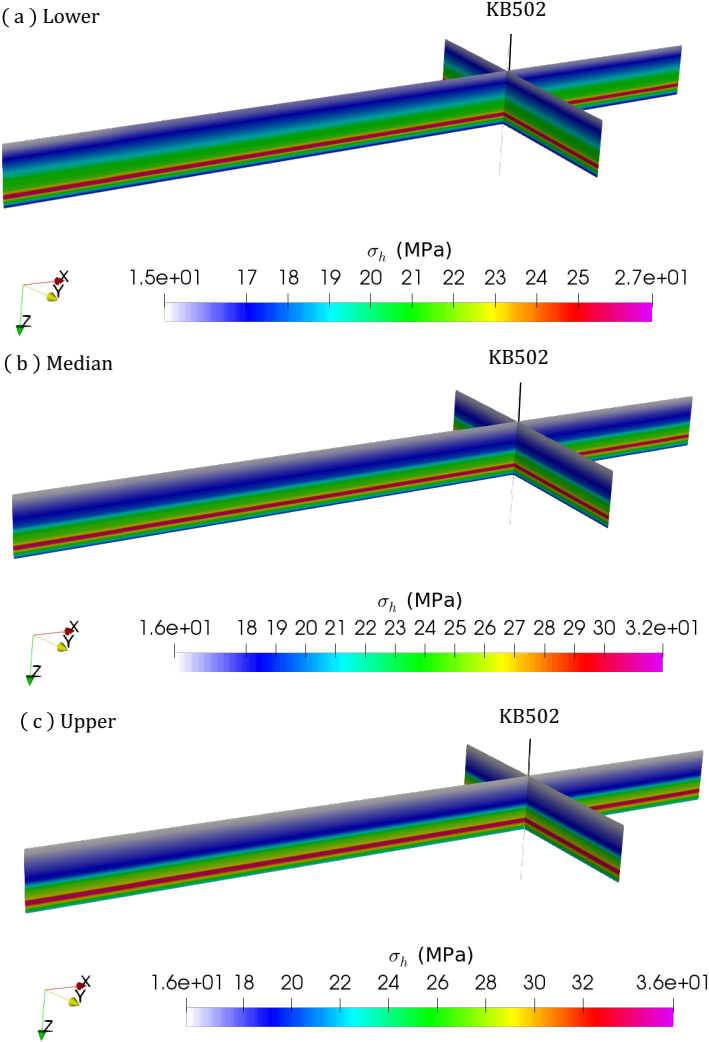

**Fig. 6.** Slices of the profile for posterior $\sigma_h$. Lower and Upper are the lower and upper of 95% CI for $\sigma_h$. Median represents the highest posterior possibility for $\sigma_h$.

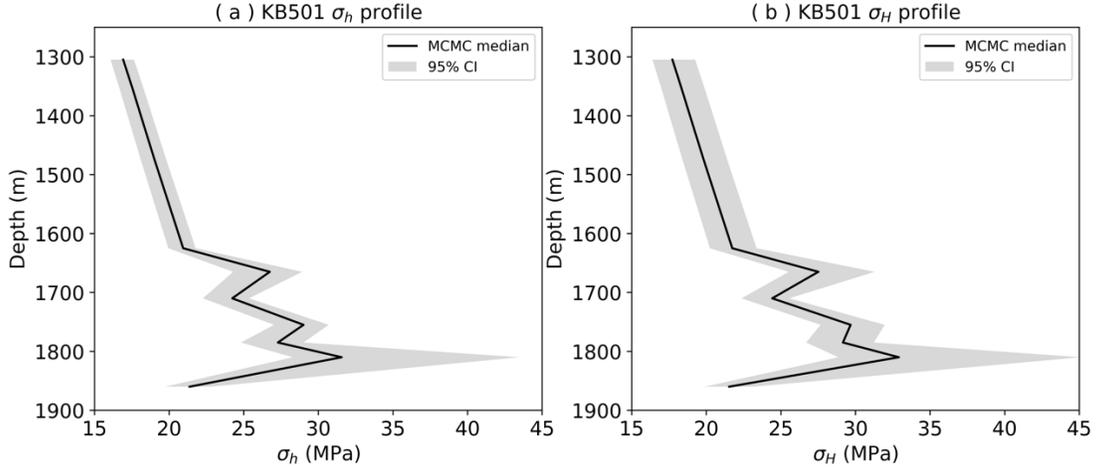

**Fig. 7.** Vertical profiles of $\sigma_h$ and $\sigma_H$ for KB-501 with quantified stress uncertainty.

## 5  Discussion

Discussions are presented in this section in terms of the 1D model throughout the following sub-sections, which will shed light on stress constraints, informative priors, and the probability of generating hydraulic fracture by injection pressures.

### 5.1  Addition of Stress Regime Constraint to Reduce Uncertainty

The scenario analyzed in Section 4.2 reveals that the posterior $\sigma_h$ and $\sigma_H$ result in the initial NF stress regime when no stress regime constraint is applied. According to Iding and Ringrose (2010), the In Salah GCS site has been inferred to be the SS stress regime, which suggests that a stress regime constraint is needed for the posterior horizontal stresses to ensure $\sigma_h \leq \sigma_v < \sigma_H$. This is discussed in this sub-section. It is worthwhile to mention that although Iding and Ringrose (2010) and some other existing studies deduced that the In Salah GCS site has a SS tendency, this may not be true for all KB-502 rock formations (Fig. 5), because given low measured in-situ $\sigma_h$ values (Table 3) and according to the stress polygon for fault failure (Zoback et al. 1987), the NF stress regime also likely exists in one or several rock formations here. Due to the lack of needed information, we assume all rock formations here are in the SS stress regime to demonstrate the Bayesian approach.

In this paper, $Q$=1.1 with a standard deviation of 0.1 is used for the SS stress regime via Eq. (9). Lecampion and Lei (2010) used the same value for their analysis of the In Salah GCS site.

This means that during the MCMC inversion analysis, in addition to matching measured $\sigma_h$, the $Q$ values at all $\sigma_h$ measurement depths are matched as well.

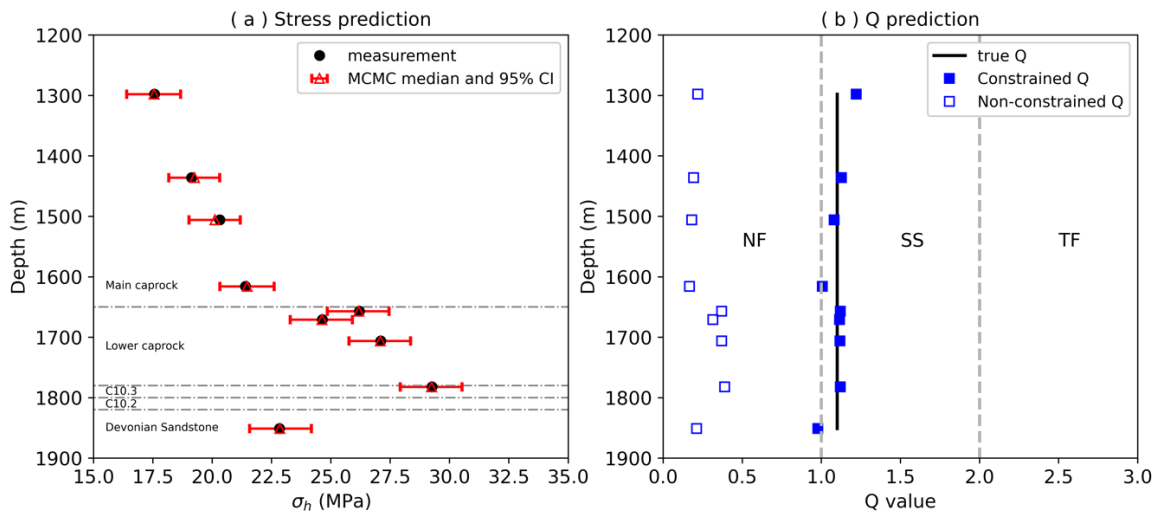

**Fig. 8.** Considering SS stress regime constraint for all measurements only (a) MCMC $\sigma_h$ predictions vs measured $\sigma_h$ data and (b) $Q$ values at all measurement depths. The bar in (a) denotes 95% CI for $\sigma_h$.

As shown in Fig. 8a, MCMC $\sigma_h$ predictions and stress uncertainty well match the measured data with no significant difference from the comparison of the no stress constraint case in Table 3. Because of the stress constraint to match the $Q$ values, the SS stress regime is found at all $\sigma_h$ measurement depths (Fig. 8b). Fig. 9 shows the posterior $\sigma_h$, $\sigma_H$, $E$, $v$, $\varepsilon_H$ and $\varepsilon_h$ under the SS stress regime constraint for KB-502. It is seen smaller uncertainty in all six parameters/variables in Fig. 9 compared to those in Fig. 5. For two horizontal stresses, 95% CI for $\sigma_h$ in the LC bottom and reservoir C10.2 has reduced ~14 MPa. 95% CI $\sigma_H$ for at all locations in Fig. 9 has even reduced very significantly and the variation of $\sigma_H$ in the 95% CI range in general is retained within 5 MPa. Thus, adding more geologic information such as the SS stress regime constraint considered here is helpful to reduce the uncertainty in stress estimations to some extent. It is worth noting that compared to $\sigma_h$, $\sigma_H$ has smaller uncertainty in the LC bottom and reservoir C10.2 (without stress measurements), primarily because both SS stress regime and frictional faulting limit constraints are applied to $\sigma_H$.

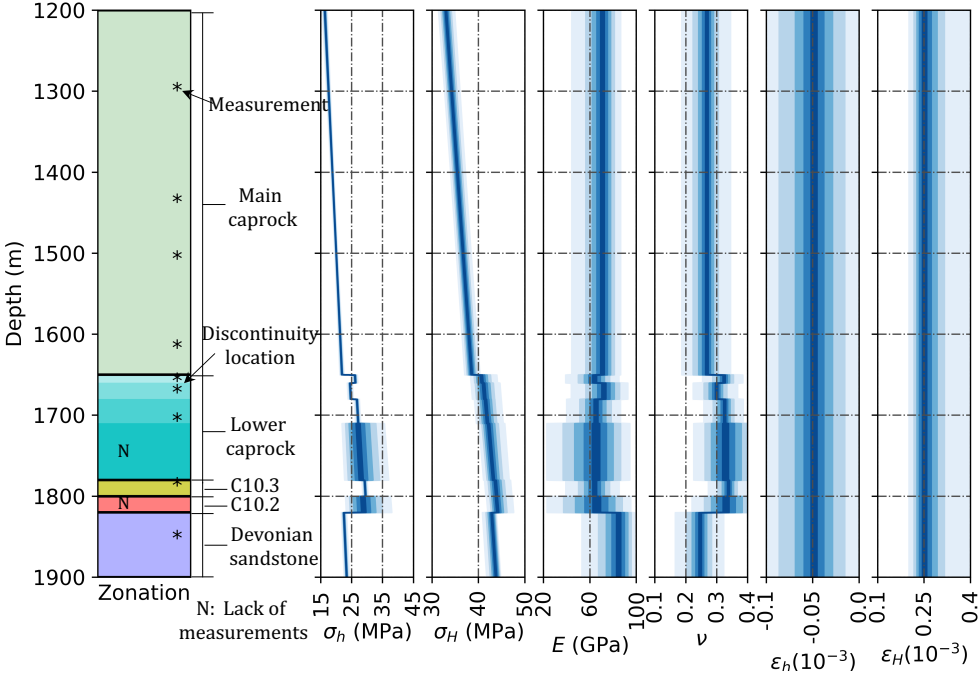

**Fig. 9.** Posterior $\sigma_h$, $\sigma_H$, $E$, $v$, $\varepsilon_H$ and $\varepsilon_h$ under the SS stress regime constraint for KB-502 along 1D depth from 1200m to 1900m.

The correlation of six parameters (as an example, including elastics parameters in the LC bottom and reservoir C10.2 without $\sigma_h$ measurements) sampled from the posterior distribution and the histograms of these parameters are plotted in Fig. 10. This plot is useful for inferring how the combination of plotted parameters has an impact on the posterior distributions. It is shown that $E$ and $v$ in the LC bottom and reservoir C10.2 have a detectable negative correlation, but the rest of the parameters have very little correlation.

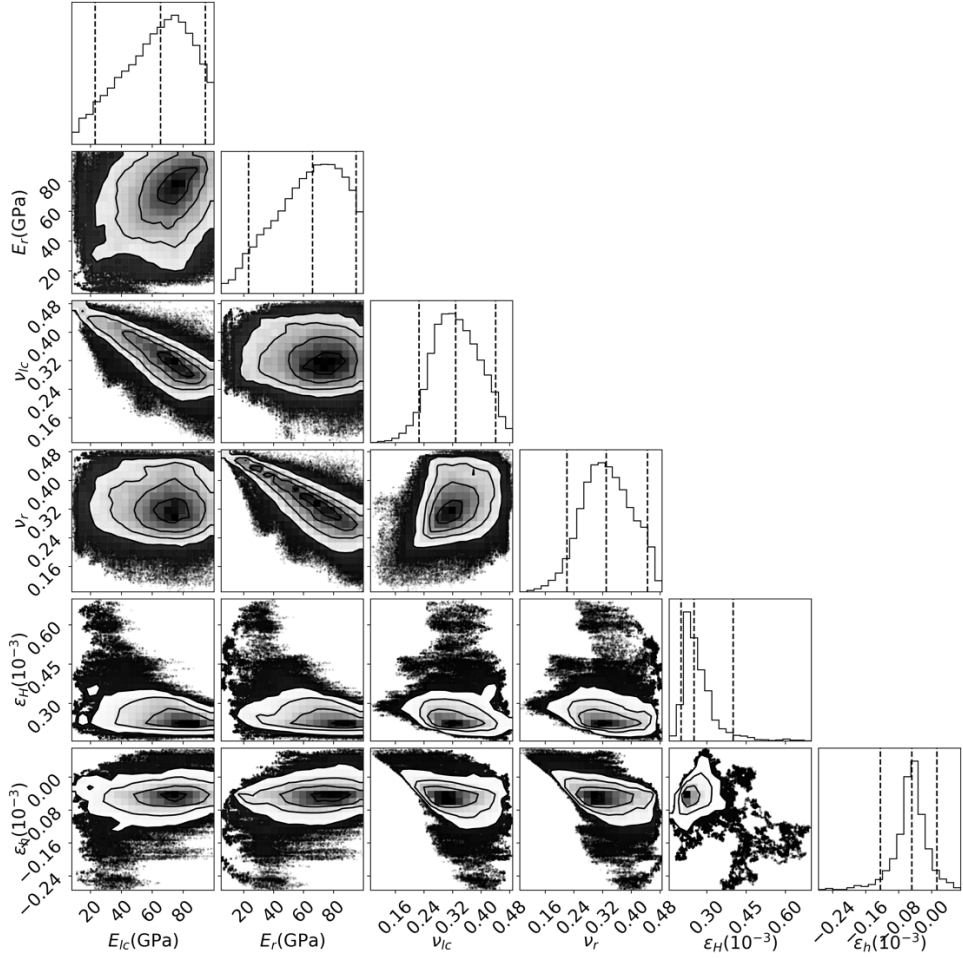

**Fig. 10.** Corner plot for the correlation of several parameters and their histograms. The subscript "*lc*" and "*r*" denotes the LC bottom and reservoir C10.2, respectively. The left and right as well as middle dashed lines in the histograms represent 95% CI and the median value.

### 5.2 Use of Information from Other Wells as Informative Priors for A Target Well

Three injection wells (i.e., KB-501, KB-502, and KB-503 in Fig. 1c) were operating separately for $CO_2$ injection in the In Salah GCS site. Table 1 shows LOT and FIT data used along KB-502 at nine depths. We have used this data for the 1D model analysis above for KB-502 and have shown that lack of stress measurements can induce significant uncertainty in $\sigma_h$ estimations. However, if we want to estimate in situ stress for another $CO_2$ injection well, e.g., KB-503 or KB-501, but this well only has a few stress measurements available, it is expected that there could be large uncertainties associated with the Bayesian stress estimations. This can be seen from Fig. 11 that shows the 1D posterior results for KB-503 under the SS stress regime. Because only one $\sigma_h$ measurement was available for the main caprock, the uncertainty in $\sigma_h$ is significant below the main caprock and the range of 95% CI for $\sigma_h$ in the lower caprock zone can be as large as 28 MPa. Therefore, finding a way to reduce such significant uncertainty is critical for the safe operation of the GCS system.

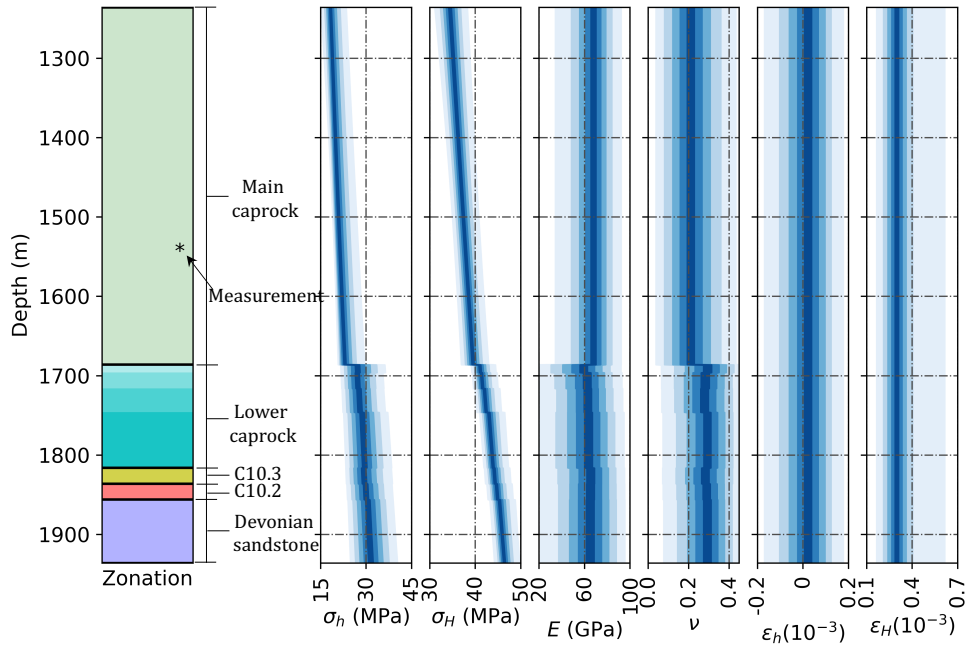

**Fig. 11.** 1D model posterior $\sigma_h$, $\sigma_H$, $E$, $v$, $\varepsilon_H$ and $\varepsilon_h$ for KB-503 under the SS stress regime constraint. The depth here for KB-503 is adjusted because the GCS site has an anticlinal structure.

One useful approach to reduce the uncertainty noted above is to locate additional information and add it as stress-related supplements into the analyzed case. This information addition approach has been recognized and demonstrated in the literature. One example comes from Feng et al. (2021), who utilized the mean stresses of five over-coring samples as informative priors to add and used to analyze the over-coring stress state of another rock sample. This approach reduced the uncertainty associated with the stress magnitude and orientation estimates. Following this idea, if the elastic parameters in each rock formation are assumed to be uniform and the stratigraphy for KB-503 and KB-502 are similar, we can construct informative priors for $E$ and $v$ in each rock formation for KB-503 using the posterior distributions for $E$ and $v$ obtained from KB-502. In other words, the mean and standard deviation of each of $E$ and $v$ from KB-502 (based on their normal distributions in Section 5.1, see Fig. 10) are used to construct the normal distributions as priors for KB-503. We test if the stress uncertainty in Fig. 11 can be reduced. Note that the tectonic strains from KB-502 are not used here.

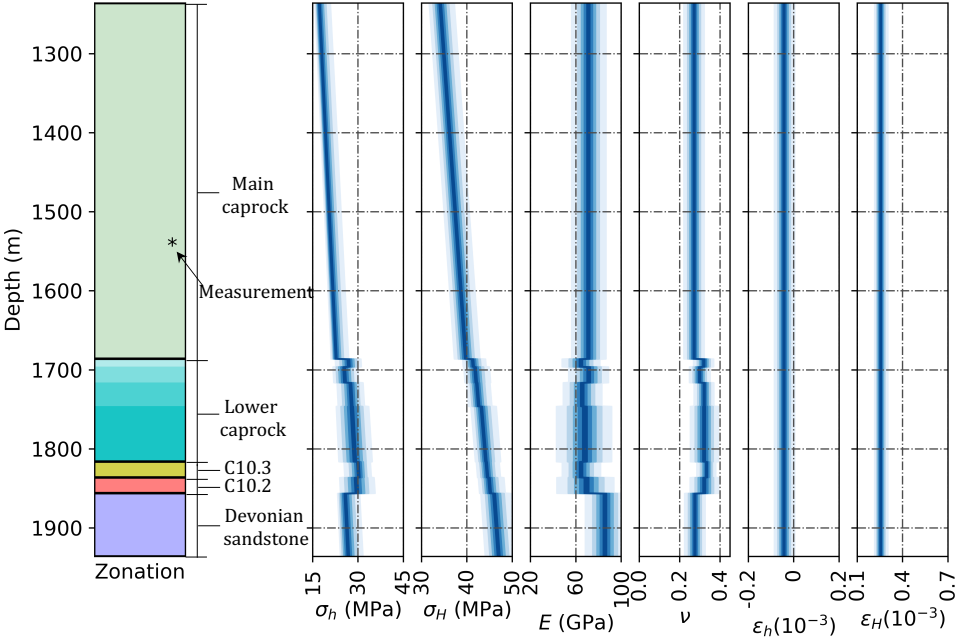

**Fig. 12.** 1D model posterior $\sigma_h$, $\sigma_H$, $E$, $v$, $\varepsilon_H$ and $\varepsilon_h$ under the SS stress regime constraint for KB-503 when informative priors are considered.

Fig. 12 plots the posterior $\sigma_h$ for KB-503 considering the informative priors. We can clearly see that the uncertainty in $\sigma_h$ is reduced significantly by adding the informative priors. In the main caprock, the 95% CI shading area for the informative priors is slightly smaller than that of the weak informative priors. However, below the main caprock where no $\sigma_h$ measurements are provided, the case of the informative prior has reduced the uncertainty in the 95% CI by as much as ~12 MPa in the lower caprock. Note that in this lower caprock unit, it is critical to have accurate estimations of the in situ stress to ensure safe $CO_2$ injection pressures that will not induce fractures or fault reactivation in the unit. The uncertainty in $\sigma_H$ is not reduced obviously compared to that in Fig. 11, but very significant reductions in uncertainty in $E$, $v$, $\varepsilon_H$ and $\varepsilon_h$ can be observed in the case of the informative prior along KB-503. The above results reveal that adding informative priors, if possible, could reduce the uncertainty in estimations of in situ stress. This is particularly helpful for many cases like the one discussed which had a limited number of stress measurements. This is also a merit of using the Bayesian inversion approach that can take advantage of both measured data and any other related information, even if imperfectly known, to obtain more reliable estimations of interest.

**5.3   Maximum Injection Pressure Recommended by Bayesian Approach**

A thorough analysis by White et al. (2014) inferred that observations of ground deformations and seismic velocity anomalies at the In Salah GCS site are very likely a result of hydraulic fractures generated in the lower portion of the caprock under injection pressures, leading to $CO_2$ migration vertically into the fractured zone. Using the Bayesian approach, we here discuss the possibility of generating hydraulic fractures in the lower caprock for KB-502 and KB-503 in terms of injection pressure data.

The injection pressure data used here is the estimated bottom hole pressure (BHP) for each well reported by Bissell et al. (2011). It is daily averaged data and estimated at a depth of ~1780 m in Tight C10.3 using measured wellhead pressures, temperatures, and flow rates. We estimate the BHP at other depths in the LC bottom and reservoir C10.2 based on the static head difference as a first order estimation. Figs. 13 and 14 show the BHP data with time at three depths, the posterior distribution for $\sigma_h$ at each depth, and the cumulative probability of generating hydraulic fractures (i.e., BHP $\geq \sigma_h$) for KB-502 and KB-503. Because the 1D modeling results for $\sigma_h$ here do not involve the effect of multiphase flow transport, the probability of fracturing here is considered as a first order estimation.

For KB-502, as shown in Fig. 13, the average BHP is ~30.7 MPa at 1780 m in C10.3 between November 2005 and August 2006 and the peak BHP is ~32.3 MPa (Fig. 13b). This average BHP results in the fracture probability nearly 100% in Tight C10.3. During the same period, the average BHP is ~30.1 MPa at 1720 m in the LC bottom where we want to estimate the possibility of generating fractures with this average BHP. It is shown that this average BHP is ~3 MPa higher than the median of $\sigma_h$ and brings the fracture possibility of ~76% under the condition that no $\sigma_h$ measurement is provided in the LC bottom (Fig. 13a).

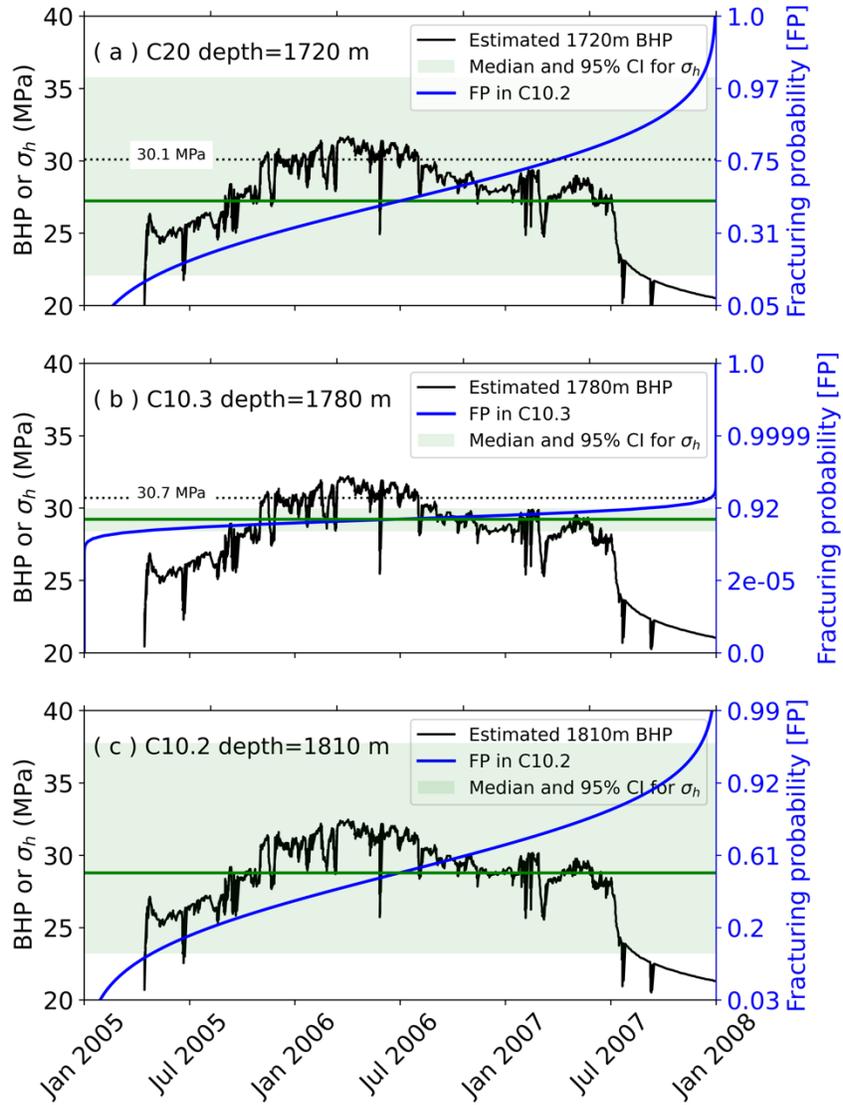

**Fig. 13.** Estimated BHP data and $\sigma_h$ 95% CI data under the SS stress regime for KB-502 at: (a) depth=1720 m in C20, (b) depth=1780 m in C10.3, and (c) depth=1810 m in C10.2. The green line is the median value for $\sigma_h$.

In the LC bottom for KB-503, as shown in Fig. 14a, the average BHP is ~27.1 MPa between March 2006 and September 2006 and ~28.1 MPa between September 2007 and December 2007. With these average BHPs, the fracture probability in the LC bottom for KB-503 is found to be ~43.4% and ~58.3% during those two periods, respectively.

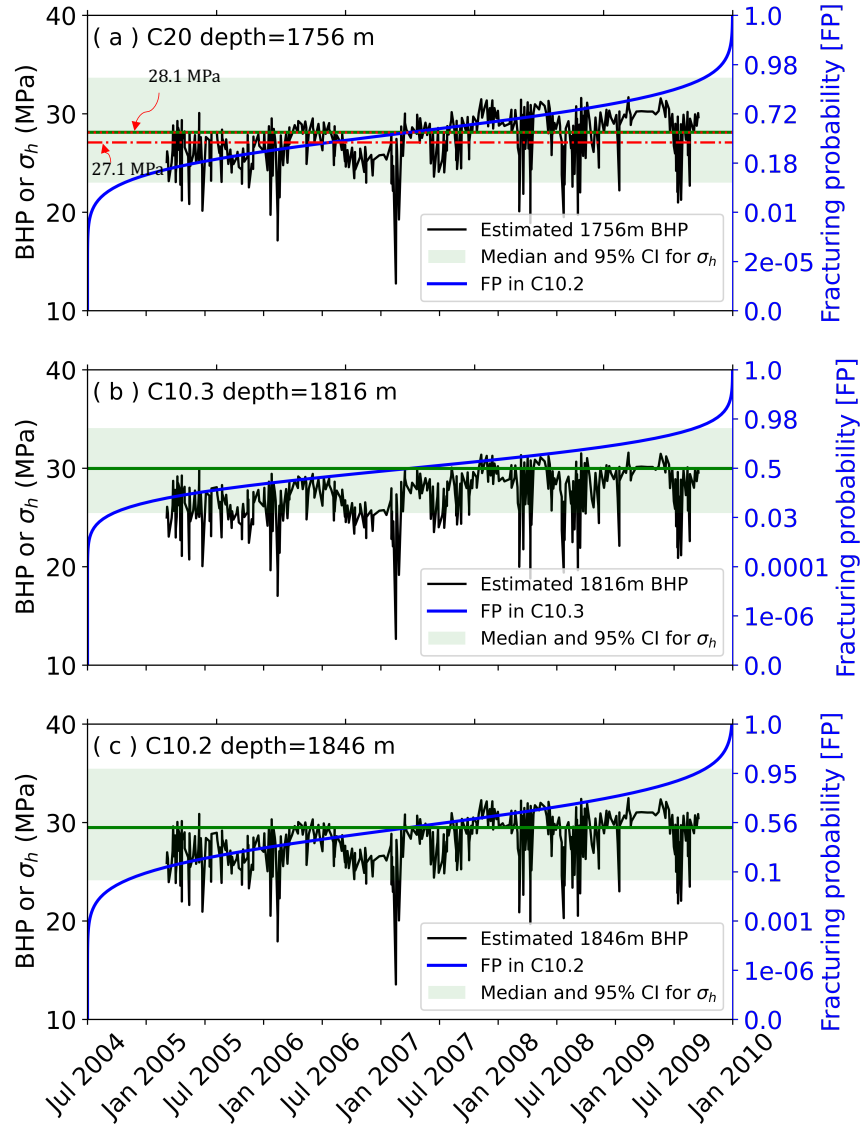

**Fig. 14.** Estimated BHP data and $\sigma_h$ 95% CI data under the SS stress regime for KB-503 at: (a) depth=1756 m in C20, (b) depth=1816 m in C10.3, and (c) depth=1846 m in C10.2. Posterior $\sigma_h$ for KB-503 is obtained from Fig. 12 with the informative priors detailed in Section 5.2. The green line is the median value for $\sigma_h$.

The analysis above shows that the probability of generating fractures in the lower portion of the caprock is nearly 100% for KB-502 and ~43.4% for KB-503 during the GCS operation time. This analysis is based on the average BHP. When the peak BHP is used for the analysis, which is treated as an extreme situation, the fracture probability increases to ~90% in the LC bottom of KB-502 (Fig. 13a). Therefore, in hindsight, the $CO_2$ operating pressure for KB-502 would have been best limited to 26 MPa or 23 MPa such that the fracture probability could be retained lower than 50% or 20%, respectively. It is worthwhile mentioning that the BHP data may also have some uncertainty due to being estimated from surface measurements only. Considering both the significant change in density and viscosity of $CO_2$ with temperature and pressure, such calculations are difficult to perform accurately. Due to the BHP uncertainty, the BHP data used above from Bissell et al. (2011) is ~3 MPa higher than the data shown in Shi et al. (2012). Even with this lower estimated BHP, the peak BHP is ~1 MPa higher than the median of $\sigma_h$ that we estimate in Fig. 13a and this could lead to the fracture probability of ~65% in the LC bottom of KB-502.

## 6 Conclusions

A Bayesian approach was proposed in this paper to provide an estimate of the in-situ state of stress throughout the volume of interest and provide an estimate of the uncertainty arising from the stress measurement, the rheology parameters, and a paucity of measurements. We have demonstrated this Bayesian approach using a data set for stress from the In Salah GCS project. Both 1D elastic-tectonic and 3D poroelastic models were analyzed in a Bayesian framework for this demonstration.

Our analysis has shown that this Bayesian approach can provide good stress estimations, but also quantify the uncertainty in stress estimations and modeling parameters, regardless of the use of the 1D or 3D models. For a case without stress measurements in the lower portion of the caprock discussed in this study, this Bayesian approach can provide the joint probability of the two horizontal principal stresses, instead of only the mean stress state of each. This allows us to make more reliable geomechanical decisions for safe GCS operation in terms of horizontal principal stresses and their uncertainty.

Two important advantages of the Bayesian approach can be highlighted in this paper. First, the uncertainty in both stress and modeling parameters can be significantly reduced by including regional geologic information (e.g., stress regime constraint) and informative priors related to the variable of interest. This reminds us to make full use of available data at the site for decision-making. Second, a maximum injection pressure can be estimated in a probabilistic manner rather than a deterministic way. With the In Salah GCS case study, we have shown that the average and peak BHPs between November 2005 and August 2006 can bring the fracture possibility of ~76% and ~90%, respectively, in the lower portion of the caprock at KB-502. Therefore, the Bayesian method in this paper can provide additional insights for site explorations and/or project operations to make informed-site decisions for subsurface engineering applications.

**Appendix**

A1. Figure supplement for 3D Bayesian modeling analysis.

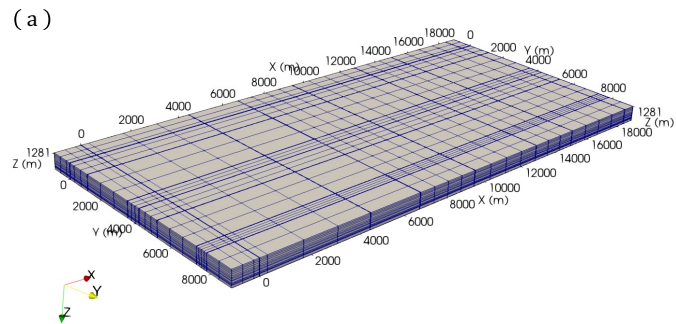

(a)

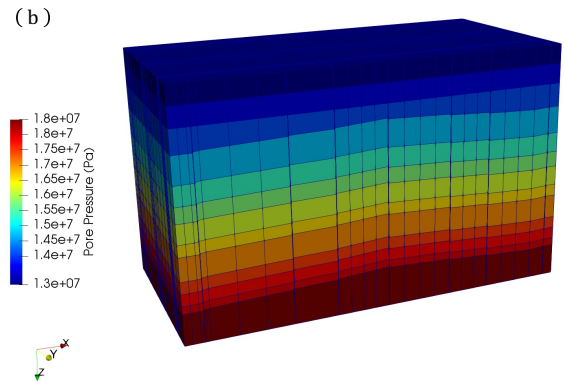

(b)

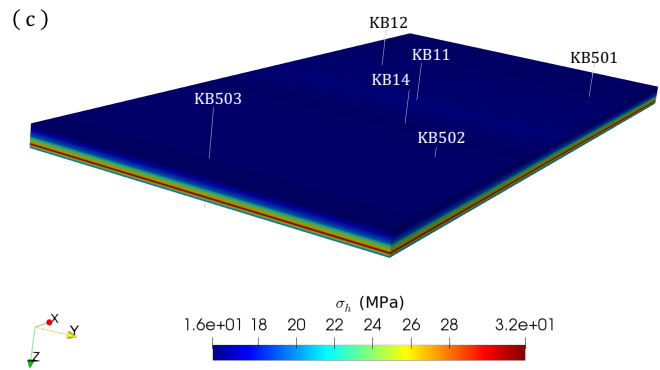

(c)

**Fig. A1.** (a) 3D model mesh. The model dimension is $20\text{km} \times 10\text{km} \times 620\text{m}$. The $x$ and $y$ axes are the direction of $\sigma_H$ and $\sigma_h$, respectively. (b) Rescaled model in the $x$ and $y$ axes (rescale factor=0.01) to show the anticlinal structure considered here. (c) 3D initial stress state for $\sigma_h$ in terms of the median of posterior $\sigma_h$. Note that the horizontal dimension is too large to visualize the anticlinal structure of the reservoir.

**Acknowledgment**

Funding for this research was provided by the National Risk Assessment Partnership (NRAP) in the U.S. DOE Office of Fossil Energy under DOE contract number DE-AC05-76RL01830. PNNL is operated by Battelle for the U.S. DOE under Contract DE-AC06-76RLO1830.

This report was prepared as an account of work sponsored by an agency of the United States Government. Neither the United States Government nor any agency thereof, nor any of their employees, makes any warranty, express or implied, or assumes any legal liability or responsibility for the accuracy, completeness, or usefulness of any information, apparatus, product, or process disclosed, or represents that its use would not infringe privately owned rights. Reference herein to any specific commercial product, process, or service by trade name, trademark, manufacturer, or otherwise does not necessarily constitute or imply its endorsement, recommendation, or favoring by the United States Government or any agency thereof. The views and opinions of authors expressed herein do not necessarily state or reflect those of the United States Government or any agency thereof.**References**

Bao T, Burghardt J, Gupta V, Edelman E, McPherson B, White M, (2021a). Experimental workflow to estimate model parameters for evaluating long term viscoelastic response of CO2 storage caprocks. International Journal of Rock Mechanics and Mining Sciences:104796.

Bao T, Burghardt J, Gupta V, White M, (2021b). Impact of time-dependent deformation on geomechanical risk for geologic carbon storage. International Journal of Rock Mechanics and Mining Sciences (accepted, in press).

Bayes T, (1763). LII. An essay towards solving a problem in the doctrine of chances. By the late Rev. Mr. Bayes, FRS communicated by Mr. Price, in a letter to John Canton, AMFR S. Philosophical transactions of the Royal Society of London:370-418.

Bissell R, Vasco D, Atbi M, Hamdani M, Okwelegbe M, Goldwater M, (2011). A full field simulation of the In Salah gas production and CO2 storage project using a coupled geo-mechanical and thermal fluid flow simulator. Energy Procedia 4:3290-3297.32